# Surface passivation of zero-mode waveguide nanostructures: benchmarking protocols and fluorescent labels

Satyajit Patra, Mikhail Baibakov, Jean-Benoît Claude, Jérôme Wenger*

*Aix Marseille Univ, CNRS, Centrale Marseille, Institut Fresnel, 13013 Marseille, France*

* Corresponding author: jerome.wenger@fresnel.fr

**Abstract**:

Zero mode waveguide (ZMW) nanoapertures efficiently confine the light down to the nanometer scale and overcome the diffraction limit in single molecule fluorescence analysis. However, unwanted adhesion of the fluorescent molecules on the ZMW surface can severely hamper the experiments. Therefore a proper surface passivation is required for ZMWs, but information is currently lacking on both the nature of the adhesion phenomenon and the optimization of the different passivation protocols. Here we monitor the influence of the fluorescent dye (Alexa Fluor 546 and 647, Atto 550 and 647N) on the non-specific adhesion of double stranded DNA molecule. We show that the nonspecific adhesion of DNA double strands onto the ZMW surface is directly mediated by the organic fluorescent dye being used, as Atto 550 and Atto 647N show a pronounced tendency to adhere to the ZMW while the Alexa Fluor 546 and 647 are remarkably free of this effect. Despite the small size of the fluorescent label, the surface charge and hydrophobicity of the dye appear to play a key role in promoting the DNA affinity for the ZMW surface. Next, different surface passivation methods (bovine serum albumin BSA, polyethylene glycol PEG, polyvinylphosphonic acid PVPA) are quantitatively benchmarked by fluorescence correlation spectroscopy to determine the most efficient approaches to prevent the adsorption of Atto 647N labeled DNA. Protocols using PVPA and PEG-silane of 1000 Da molar mass are found to drastically avoid the non-specific adsorption into ZMWs. Optimizing both the choice of the fluorescent dye and the surface passivation protocol are highly significant to expand the use of ZMWs for single molecule fluorescence applications.

The diffraction of light ultimately limits the optical performance of confocal microscopes and restricts their ability to interrogate a single molecule in a crowded environment.[1–3] To overcome the diffraction limit, zero-mode waveguides (ZMWs) have been introduced as efficient means to confine the light at the nanometer scale.[4,5] ZMWs are nanoapertures of 50 to 200 nm diameter milled in opaque metallic films. Thanks to their subwavelength diameter, ZMWs generate an evanescently decaying intensity profile,[6] which offers an effective detection volume in the attoliter ($10^{-18}$ L) range, three orders of magnitude below the diffraction-limited confocal volumes.[1,4] Another major advantage of ZMWs concerns their ability to enhance the fluorescence brightness per molecule thanks to the enhanced local excitation intensity inside the ZMW and the modification of the fluorescence photokinetics decay rates.[7–16]

Since their inception in 2002, ZMWs have been widely used for a large range of biophysical and biochemical applications, including DNA sequencing,[17–19] enzymatic reaction monitoring,[20–22] protein-protein interaction,[23–27] nanopore sensing,[28–32] Förster resonance energy transfer,[33–35] biomembrane investigations,[36–40] and nano-optical trapping.[41–44] However, all these applications require a proper surface passivation and/or functionalization of the ZMW in order to avoid the unwanted adsorption of the fluorescent molecules onto the ZMW surface that would impede the experiments. While extensive literature exists for passivating glass surfaces for single molecule fluorescence microscopy,[45–52] little is known about the surface passivation of ZMWs and the comparative analysis of different passivation strategies. For aluminum ZMWs, most studies reproduce the protocol based on polyvinylphosphonic acid (PVPA) as initially introduced by Korlach and coworkers in Ref. [53]. For gold ZMWs, a thiol-derivatized polyethylene glycol (PEG) has been reported.[54] However, the nature of the molecular adhesion phenomenon remains unclear in the ZMW. Also, the performance of different surface passivation approaches remains to be quantitatively compared in a clear benchmark study. Moreover, ZMWs generally use aluminum films to operate in the blue-green region of the visible spectrum, but aluminum is quite unstable in water environments and can be corroded within a few days or less.[55–57] Recently, it was shown that the surface passivation of the aluminum layer can greatly improve the chemical stability in water buffers,[58–60] which points out another advantage for the surface passivation of aluminum ZMWs.

Here, we first explore the nonspecific adhesion of fluorescently-labelled double-stranded DNA molecules on the surface of aluminum ZMWs. We show that the nonspecific surface adhesion of DNA is directly correlated with the choice of the fluorescent dye label, its surface charge and its hydrophobicity. Specific guidelines are discussed to properly select the fluorescent label,[61–64] as our data indicates that negatively-charged hydrophilic dyes have negligible affinity for the ZMW surface while positively-charged moderately-hydrophobic dyes clearly promote the surface adhesion of DNA double strands in our experimental conditions. Having identified the conditions that lead to a clear molecular affinity for untreated ZMW surfaces, we then benchmark various surface passivation protocols. The most widely used approaches using bovine serum albumin (BSA), polyethylene glycol (PEG) of various chain lengths and polyvinylphosphonic acid (PVPA) are quantitatively compared. Dichlorodimethylsilane (DDS) is another popular approach to passivate glass structures,[48] yet it is highly corrosive for aluminum and quickly dissolved the ZMW structures. Therefore, DDS was excluded from our study. Importantly, we demonstrate that quantitative measurements using fluorescence correlation spectroscopy (FCS) or fluorescence lifetime can still be reliably performed inside ZMWs after proper surface passivation. Highlighting the role of the fluorescent label and the adequate surface treatment provides important knowledge to further expand and ease the use of ZMWs for single molecule fluorescence applications.

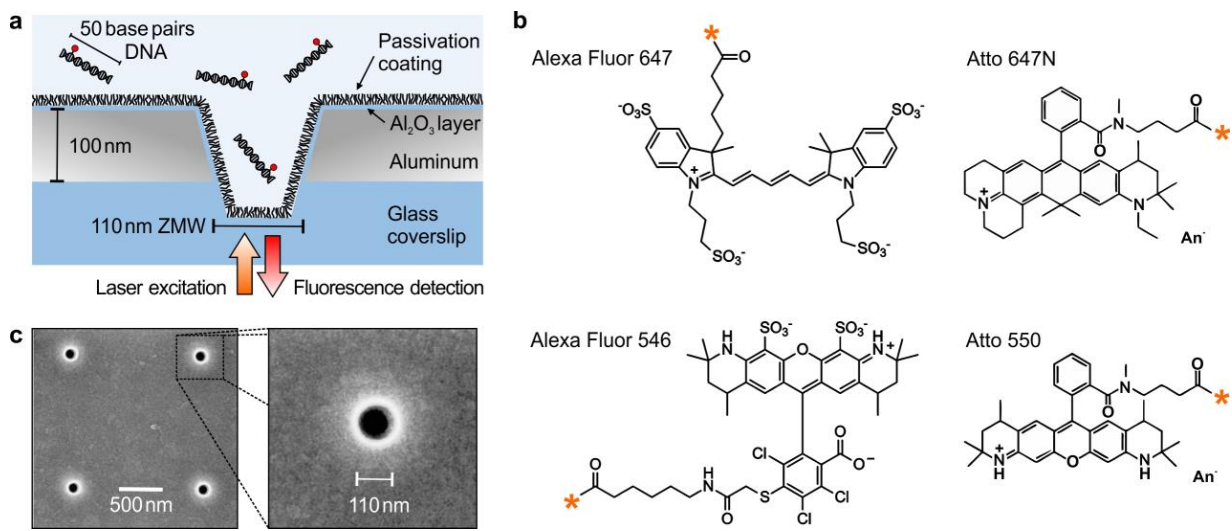

**Figure 1**. Aluminum zero-mode waveguide nanoaperture (ZMW) and fluorescent dye labels used to investigate the surface passivation. (a) Schematic of our experiment with a single 110 nm aluminum ZMW on a glass coverslip. The ZMW is covered with a solution of fluorescent labelled DNA which diffuse across the ZMW. (b) Chemical structures of Atto and Alexa dye molecules used to label the DNA. The labeling sites are indicated by the orange star. (c) Scanning electron microscope (SEM) images of aluminum ZMWs.

**Results**

**Fluorescent dye induced adhesion of DNA molecules on the ZMW surface.** Figure 1a shows a scheme of our experiment intended to investigate the effects of unspecific adsorption of dye-labelled double stranded DNA molecules on the ZMW surface. The ZMWs are fabricated on a 100 nm thick aluminum film deposited on a glass microscope coverslip (Fig. 1c). Here, we focus on a fixed ZMW diameter of 110 nm as this size was shown to provide near-optimal fluorescence enhancement performance for both green and red dyes.[35] We observed similar adsorption effects for ZMW of diameters in the 85-200 nm range. So, our findings and conclusions reported here are quite general to ZMWs and can be applied other aluminum plasmonic nanostructures as well.

The chemical structures of the Atto and Alexa dyes used to fluorescently label the DNA sample are shown in Figure 1b. Both Atto and Alexa dyes have high absorption cross sections and fluorescence quantum yields, and are amongst the most popular and extensively used fluorescent labels in single molecule fluorescence measurements. Here, we have selected both green (Atto 550 and Alexa 546) and red (Atto 647N and Alexa 647) dyes. Only one fluorescent dye is present on each double stranded DNA, and HPLC purified samples are used to clearly investigate the influence of the dye chosen. Importantly, the 110 nm aluminum ZMW features a very similar optical performance in both the green and red spectral windows,[35] so the data for green and red dyes can be readily compared. The ZMW sample is covered with a solution of dye labelled DNA which can diffuse across the ZMW. With the optical configuration displayed in Fig. 1a, only the fluorescence signal stemming from the ZMW attoliter volume is collected. Figure 2a-c display the fluorescence intensity time trace and its temporal correlation function for Alexa 647 and Atto 647N-DNA conjugates inside passivated and unpassivated ZMWs. Similar raw fluorescence data are shown for Alexa 546 and Atto 550-DNA in the Supporting Information Fig. S1.

For Alexa 647-DNA conjugates, a flat intensity time trace is observed inside non-passivated ZMWs (Figure 2a). Short duration peaks are visible on the trace, and correspond to the standard Poisson noise for fluorescent species freely diffusing across the ZMW volume. Importantly, the trace for Alexa 647-DNA is free of longer duration spikes. These features are confirmed by the fluorescence correlation spectroscopy (FCS) analysis of the time trace,[65] which shows only a single fast diffusion component (Figure 2a, right). The situation is clearly different when the same experiment is redone with Atto 647N-DNA conjugates instead of Alexa 647-DNA (Figure 2b). The fluorescence time trace for Atto 647N-DNA conjugates exhibits several major spikes of duration longer than 100 ms. The corresponding FCS correlation function confirms the trend with a supplementary component at long lag times. To confirm that the observed long fluorescence spikes are related to the adhesion of Atto 647N-DNA on the ZMW surface, we passivate the ZMW with a silane-modified polyethylene glycol of 1000 Da molecular weight (PEG 1000). The ZMW passivation with PEG 1000 completely removes the contribution from long duration events for Atto 647N labelled DNA (Figure 2c). For PEG 1000 passivated ZMWs, we retrieve the same features for Atto 647N-DNA than for Alexa 647-DNA, as Figures 2a and 2c are very similar and correspond to the awaited situation of free diffusion without any significant molecular affinity for the surface. Altogether, these results demonstrate that the long fluorescence spikes seen in Figure 2b are related to the adhesion of the Atto 647N labelled DNA on the ZMW surface. Despite the small size of the dye as compared to the 51 base pair DNA, the choice of the fluorescent dye clearly plays a crucial role in the unwanted adhesion to the ZMW surface: Atto 647N labelled DNA bears a strong affinity to adhere on the surface, whereas Alexa 647-DNA does not. Similar observations were performed for the green dyes (Supporting Information Fig. S1). Alexa 546-DNA does not show significant affinity for the surface while Atto 550-DNA presents clear signs of surface sticking.

Already the data in Fig. 2a-c demonstrate that the external fluorescent marker plays a key role in promoting adhesion of the DNA on the ZMW surface. The Atto 647N and Atto 550 dyes appear more prone to inducing sticking than the Alexa 647 and Alexa 546 dyes. Their chemical structures (Figure 1b) indicate that the Alexa dyes bear a negative charge and have more charged group while the Atto dyes have a positive charge after DNA labelling.[63,64] Therefore, these Atto dyes seem to be comparatively more hydrophobic than their Alexa counterparts. Hydrophobicity can be described quantitatively with the distribution coefficient or partition coefficient represented by logD where D denotes the ratio of the concentration of the solute between a nonpolar (such as octanol) and a polar (such as water) solvent. So, a solute with a positive value of logD is hydrophobic and a solute with negative value of logD is hydrophilic. For Atto 550 and Atto 647N the logD value is estimated as 6.41 and 3.26 respectively, while for Alexa 546 and Alexa 647 the value of logD is found to be -1.43 and -4.26 respectively.[64] The more positive value of logD for the Atto dyes clearly indicates that the Atto dyes are more hydrophobic in nature in comparison to Alexa dyes. Our results indicate that the dye hydrophobicity is a crucial parameter to promote surface adhesion of labelled DNA molecules. Similar effects were observed for glass slides while comparing proteins labelled with cyanine 5 and Atto 647N,[48] and while observing dye-induced binding of proteins onto glass surfaces,[63] or lipid bilayers.[64] Our results on DNA inside aluminum ZMWs show that the surface affinity trend is quite general and that the physicochemical properties of the organic fluorescent dyes should be considered carefully while designing the experiment.

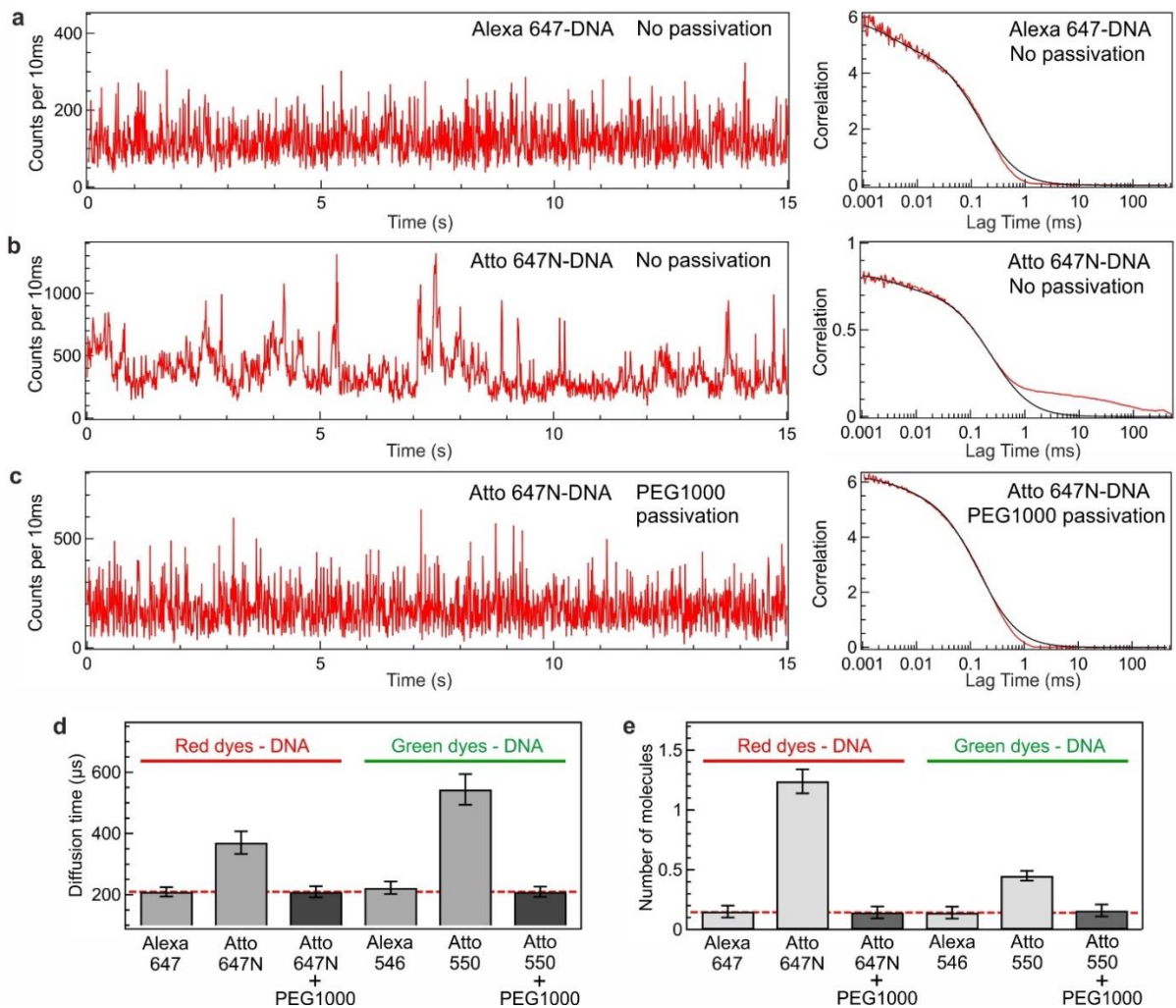

**Figure 2**. Fluorescent dye induced adhesion of DNA molecules on the ZMW surface. (a-c) Fluorescence intensity time trace and corresponding FCS correlation function for (a) Alexa 647-DNA on a ZMW without any surface passivation, (b) Atto 647-DNA on a ZMW without any surface passivation and (c) Atto 647-DNA on ZMW with surface passivation using PEG1000. On trace (b), longer time duration fluorescence peaks (spikes) and a longer diffusion component in the FCS trace indicate the unspecific adsorption of the labelled DNA molecules on the ZMW surface. FCS analysis quantifies the diffusion time (d) and the number of detected molecules (e) for the different DNA labels and in the presence or absence of additional PEG1000 surface passivation. The DNA concentration is constant at 100 nM and the ZMW diameter is 110 nm. The dashed red lines in (d,e) indicate the expected results in the absence of sticking events (purely free diffusion).

We go one step further and show that the surface sticking affects the quantitative results from FCS analysis by introducing major artefacts. The FCS traces in Figure 2a-c and S1 are fitted to determine the number of molecules (N) and diffusion time ($\tau_D$) of the dye-DNA conjugates inside the ZMW (in the case of Atto 647N and Atto 550 labelled DNA in uncoated ZMW the longer diffusing component are excluded from the fit in order to determine the N and $\tau_D$). Normally, all results should converge towards similar values as we use the same DNA concentration and same ZMW diameter. However, the results for Atto 647N and Atto 550-DNA inside non-passivated ZMWs clearly deviate from the expected trend (Figure 2d,e). In the occurrence of surface sticking, the parameters measured by FCS can be drastically different than what is observed for freely diffusing molecules. The relative error for the diffusion time can be greater than 2×, while the number of molecules can be misestimated by 10×! Similarly, the fluorescence brightness per molecule is also affected (Supporting Information Fig. S2). These large deviations indicate that proper care must be taken to choose the fluorescent label wisely or passivate the surface effectively to avoid observing the deleterious effects of sticking events. Alternatively, the results in Figure 2d,e also show that quantitative FCS measurements remain possible inside ZMWs with proper surface passivation.

While FCS is known to be very sensitive to any fluctuation affecting the fluorescence intensity, the fluorescence lifetime is comparatively believed to be much more robust to fluctuations or misalignments as it only relies on the arrival time of the fluorescence photons. However, Figure 3 shows that the dye-induced adhesion to the ZMW surface also affects the recorded fluorescence decay dynamics and introduces artefacts in the measured fluorescence lifetimes. Fluorescence decay traces of Alexa 647 and Atto 647N labelled DNA in confocal and inside a 110 nm ZMW are shown in Figure 3a and 3b (decay traces for the green dye-DNA conjugates are shown in the Supporting Information Figure S2). For all the dye-DNA conjugates, the fluorescence decays inside the ZMWs are faster than the confocal reference. This indicates a reduction of the fluorescence lifetime due to the modified electromagnetic environment inside the ZMW.[7,8] However, looking into more details in the occurrence of surface adhesion (Atto 647N and Atto 550-DNA conjugates inside unpassivated ZMW), we observe that the fluorescence lifetime is significantly shortened as compared to the non-sticking cases. We quantify the lifetime reduction (and the influence of sticking) by computing the ratio of the fluorescence lifetime between the confocal reference ($\tau^0$) and ZMW ($\tau^{ZMW}$) for the different dye-DNA conjugates (Figure 3c). For Alexa dyes-DNA the ratio of $\tau^0/\tau^{ZMW}$ is similar for passivated and unpassivated ZMWs and is about 2-fold, confirming that no significant surface adhesion is observed with these dyes. However, for Atto-DNA constructs, a larger 3-fold lifetime decrease is observed inside unpassivated nanoapertures where sticking occurs (Figure 3c). Hence not only are quantitative FCS measurements affected by the dye-induced DNA sticking, but also the fluorescence lifetime can be significantly shifted. The further decrease in the fluorescence lifetime in the case of sticking indicates that the dye molecules preferentially adsorb on the metal surface which opens a supplementary nonradiative deactivation route through energy transfer from the Atto dye to the free electrons of the metal. Again, we find that surface passivation of the ZMW with PEG 1000 efficiently recovers the expected fluorescence decay dynamics for the Atto-DNA conjugates, and avoids the negative effects of surface adhesion.

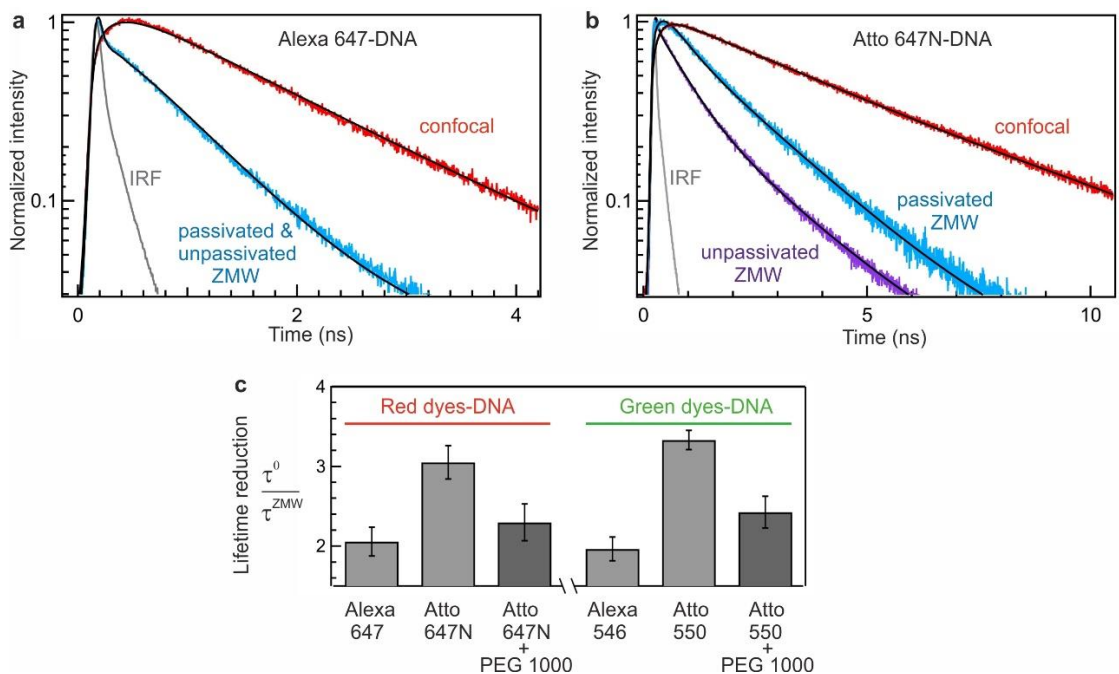

**Figure 3**. Fluorescent dye induced adhesion of DNA molecules also affects the observed fluorescence lifetime dynamics. (a) Fluorescence decay trace of Alexa 647-DNA and (b) Atto 647N-DNA conjugates in confocal and in a 110 nm ZMW. Black lines are the reconvoluted fits to the data by taking into account the instrument response function (IRF). In the case of Alexa 647 the intensity decay is identical for unpassivated and passivated ZMWs. However, in the case of Atto 647N the intensity decay is significantly faster for unpassivated ZMW than the passivated one. (c) Ratio of fluorescence lifetime between confocal reference ($\tau^0$) and ZMW ($\tau^{ZMW}$) for labelled DNA with the green (Alexa 546 and Atto 550) and red dyes (Alexa 647 and Atto 647N) inside passivated and unpassivated ZMW. In the case of Alexa dyes, the results are identical for unpassivated and passivated ZMW.

**Surface passivation approaches to prevent the adhesion of Atto 647N-DNA.** So far, we have demonstrated that despite the small size of the fluorescent label as compared to the 51 base pairs DNA molecule, the choice of the dye and its surface charge after labelling plays a key role in promoting the adhesion of DNA molecules in the ZMW surface. Incidentally, our data also shows that surface passivation using the PEG 1000 protocol is an efficient way to avoid the surface adhesion in the case of Atto 647N and Atto 550 dyes. In the following, we explore the relevance of different surface passivation approaches and compare their efficiencies. We investigate the most commonly used surface passivating agents like bovine serum albumin (BSA), polyethylene glycol (PEG) of different molecular weights (PEG 500, PEG 1000 and PEG 5000) and polyvinyl phosphonic acid (PVPA). Dichlorodimethylsilane (DDS) is also often used as a glass surface passivating agent.[46–48,52] However, DDS is highly corrosive for aluminum structures and could not be used with our Al ZMWs. Despite the presence of the alumina layer covering the aluminum film, we found that DDS dissolved the aluminum film within a few minutes as a consequence of the chloride reaction with aluminum. Therefore, DDS was excluded from our test.

Our experiments focus on Atto 647N-labelled DNA as this dye is most prone to induce surface sticking. Figure 4 shows the results obtained with the different ZMW surface passivations. BSA and PEG 5000 passivations do not lead to a significant improvement as compared to the results for an uncoated ZMW (Figure 4a, b). Long duration spikes are observed on the intensity time traces and the FCS correlations clearly exhibit a longer diffusion component. These results indicate that the BSA and PEG 5000 surface passivations are insufficient to fully prevent the Atto 647N-DNA adhesion. Similar results are also observed for PEG 500 passivated ZMWs (Supporting information Fig S4). However, a clear improvement is observed for PEG 1000 and PVPA passivations as the long duration spikes are efficiently removed from the fluorescence traces (Figure 4c, d). The resulting FCS traces also do not display any longer diffusion component for PEG 1000 and PVPA passivation, confirming the removal of the surface adhesion events. In particular, the PVPA passivation helps to further clarify the origin of the surface adhesion coming from the metallic cladding or silica bottom of the ZMW. As PVPA was reported to selectively bind to the metal surface,[53] it appears that the metal cladding of the ZMW is mainly responsible for the DNA sticking. The shortening of the lifetime of the Atto dyes inside the unpassivated ZMW further confirms this interpretation (Figure 3b).

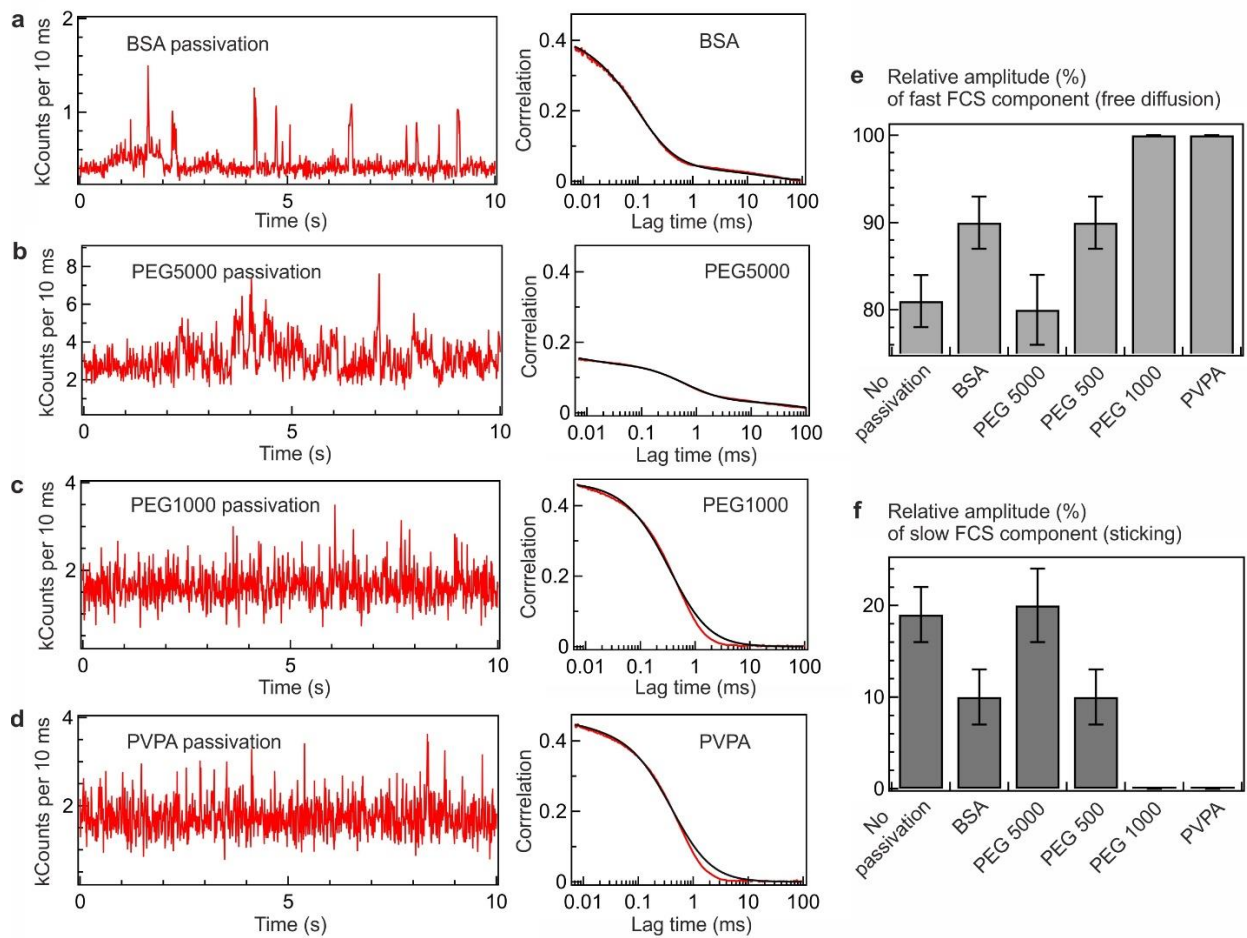

**Figure 4**. Different surface passivation approaches to prevent the adhesion of Atto 647N-DNA. (a-d) Fluorescence time traces and FCS correlation functions of the Atto 647N-DNA conjugate in the 110 nm aluminum ZMW passivated with (a) BSA, (b) PEG 5000, (c) PEG 1000 and (d) PVPA. The FCS traces were fitted using a two species model to determine the amplitude of the short and long diffusion components. (e,f) Relative amplitudes of the fast (e) and slow (f) diffusion components for different ZMW surface passivations.

For a quantitative comparison of the passivation performance between the different protocols, we have fitted the FCS traces in Fig. 4a-d with a two component diffusion model to measure the amplitude of the longer diffusion component indicative of sticking. The relative amplitude of the slow diffusion component basically reflects the amount of sticking fraction of the DNA, while the relative amplitude of the fast component indicates the percentage of the diffusion limited events. In the absence of sticking, we therefore expect that the relative amplitude of the fast component approaches 100% while the amplitude of the long component vanishes (Figure 4e,f). In the case of unpassivated ZMWs, the long component amplitude is around 20 %. Passivation with PEG 5000 reduces the correlation time of the long component related to sticking/unsticking (Supporting Information Fig. S4), yet it does not yield any significant improvement in reducing the amplitude of the long component, as the PEG chain length is too long and PEG surface density is too low to efficiently inhibit the Atto 647N-DNA surface adhesion. Using a shorter PEG chain of 500 Da (PEG 500) improves the situation by reducing the long component relative amplitude to 10%, but PEG 1000 (or PVPA) provides a much better optimum as in this case the long component is nearly absent from the FCS trace. Additionally, the widely used BSA passivation approach is not very efficient here, as a 10% fraction of long component amplitude could still be detected in this case.

**Conclusion**

Zero mode waveguides provide a simple and efficient mean to overcome the diffraction limit in single molecule fluorescence detection by enabling single molecule analysis at micromolar concentrations with enhanced fluorescence brightness. However, the use of the ZMW structures for quantitative fluorescence measurements can be severely hampered by unwanted surface adhesion events affecting both FCS and fluorescence lifetime results. Two main conclusions can be driven from our present work. First, our data demonstrates that the choice of the external fluorescence label plays a key role in promoting the surface adhesion of DNA double strands. In our experimental conditions, positively-charged moderately-hydrophobic Atto 550 and Atto 647N dyes tend to promote surface sticking of DNA, while negatively-charged hydrophilic Alexa 546 and Alexa 647-DNA conjugates do not show any detectable affinity for the surface. While working with standard aluminum surfaces and buffers (20 mM Hepes, 10 mM NaCl at pH 7.5), the choice of a fluorescent label with a net negative charge and good hydrophilicity appears to be important to minimize observing unwanted surface adhesion events of DNA double strands. Second, different surface passivation approaches are quantitatively benchmarked to determine the most efficient approach. Even in the presence of a high level of sticking such as for the Atto dyes used here, the ZMWs can be still used without introducing artefacts if their surfaces are adequately passivated. While the commonly used approaches of BSA and PEG 5000 are not efficient for ZMWs, the two options using PEG 1000 and PVPA efficiently remove all detectable adhesion events. These findings are highly relevant to enable the wide use of ZMWs and aluminum nanophotonics for single molecule biophysics and biochemistry studies.

**Methods**

**Zero-mode waveguide preparation.** A 100 nm thick layer of aluminum is deposited on a clean borosilicate glass microscope coverslip.[35] The metal deposition is performed by electron-beam evaporation (Bühler Syrus Pro 710), with a deposition rate of 10 nm/s at a chamber pressure of 5. $10^{-7}$ mbar. The ZMWs are then carved into the aluminum layer using a gallium-based focused ion beam (FEI dual beam DB 235 Strata). The focused ion beam parameters are set to 30 kV voltage and 10 pA current, and the gallium ion beam has a resolution of about 10 nm.

**DNA Sample.** A fluorescently labelled double stranded DNA of 51 base pairs length is designed for the present study. Both green (Atto 550 and Alexa 546) and red (Atto 647N and Alexa 647) dyes are used as fluorescent labels. The forward strand sequence of the DNA is 5'-CCT GAG CGT ACT GCA GGA TAG CCT ATC GCG TGT CAT ATG CTG T**T**C AGT GCG-3' where the thymine at position 44 is labelled with a green dye (Atto 550 and Alexa 546) to prepare the green labelled construct. The sequence of the complementary reverse strand of the DNA is 5'-CGC ACT GAA CAG CAT ATG ACA CGC GAT AGG CTA TCC TGC AGT ACG C**T**C AGG-3' where the T base at position 47 is labelled with a red dye (Atto 647N or Alexa 647) for the preparation of red labelled DNA. Please note that only one fluorophore is present in the double stranded DNA used in the measurements, as the complementary strand for each case does not contain any fluorescent label.

All the HPLC purified DNA strands are obtained from IBA life solution (Göttingen, Germany). The appropriate forward and reverse stands of the DNA are annealed according to a reported procedure to prepare the double stranded DNA. First, the forward and its complementary reverse strand were mixed in 1:1 molar ratio in a hybridization buffer containing 5 mM Tris, 20 mM $MgCl_2$, 5 mM NaCl at pH 7.5 to prepare the 5 µM double stranded DNA. First, the mixture is heated rapidly at 90°C for 5 minutes. After then the mixture is slowly and gradually cooled down to room temperature for over 3 hours. The annealed double stranded DNA is diluted in 20 mM Hepes, 10 mM NaCl, 0.1% Tween 20, pH 7.5 buffer for the measurements. In the first set of measurement (Fig. 2) both Alexa and Atto labelled DNA constructs are used at 100 nM to study their fluorescence behavior in presence of unpassivated and PEG passivated ZMW. In the second set of experiments (Fig. 4), around 400 nM of Atto 647N-DNA conjugate is used to study the effect of different surface passivating agents in eliminating the sticking of the DNA on the ZMW surface. Tris(hydroxymethyl)aminomethane (Tris, ≥99.8%), Hepes (≥ 99.5 %, molecular biology grade), NaCl, $MgCl_2$ and Tween 20 are purchased from Sigma Aldrich and used without further purification.

**Surface passivation protocols.** Silane-modified polyethylene glycol (PEG) of molecular weight 500 Da (PEG 500) and 1000 Da (PEG 1000) are purchased from Interchim. First, the aluminum nanoapertures are rinsed with water, ethanol and isopropanol and then treated with air plasma for 5 minutes to remove any organic impurities from the nanoapertures. Immediately after cleaning, the nanoapertures are covered with a solution of 1 mg/ml PEG-silane in absolute ethanol (≥ 99.7%) with 1% acetic acid (AR grade) for overnight at room temperature (20°C) under Ar environment. The nanoapertures are then washed with anhydrous ethanol to remove any unadsorbed PEG-silane and dried in a flow of synthetic air.

For the surface passivation with PEG of molecular weight 5000 Da (PEG 5000) a two-step protocol is used following a reported procedure. In the first step, the aluminum ZMW surface is silanized by placing the cleaned ZMW sample in a solution of 1% (V/V) 3-aminopropyltriethoxy silane in slightly acidic (5 % V/V acetic acid) methanol for 15 minutes. After then, the excess silane is washed away by rinsing with MilliQ water and then dried in a stream of synthetic air. In the second step, the silanized ZMWs are incubated with the solution of methoxy polyethylene glycol succinimidyl carboxymethyl (PEG 5000 NHS ester, Iris Biotech GmbH PEG1165.0001) in 100 mM $NaHCO_3$ buffer (pH 8.25) for 3-4 hours to prepare the pegylated nanoaperture surface where the PEG 5000 chain are covalently linked to the aminosilanes via the NHS ester group reaction. After the incubation, the nanoapertures are repeatedly rinsed with MilliQ water to remove any excess unattached PEG and dried by using a stream of synthetic air.

Poly(vinyl)phosphonic acid (PVPA) is used to selectively passivate the metal surface.[53] For PVPA passivation, the cleaned aluminum ZMWs are covered with an aqueous solution of 2.8% m/V PVPA and then heated at 90°C for 15 minutes. Excess PVPA is washed away by rinsing with MilliQ water. The ZMWs are then dried by using a flow of synthetic air and finally annealed at $80^0$C for 10 minutes in dry atmosphere.

To passivate with bovine serum albumin (BSA) protein the cleaned aluminum ZMWs are covered with a solution of 1 mg/ml BSA in a buffer containing 10 mM Tris HCl, 10 mM NaCl at pH 8.0 for 30 minutes. The excess unadsorbed BSA is removed by washing the ZMWs with MilliQ water.

**Experimental setup**. The fluorescence measurements are carried out in a custom build confocal microscope set up, as described in Ref. [35]. The green dyes (Atto 550 and Alexa 546) are excited at 557 nm by a iChrome-TVIS laser (Toptica GmbH, pulse duration ~ 3 ps). A LDH series laser diode (PicoQuant, pulse duration ~ 50 ps) is used to excite the red dyes (Atto 647N and Alexa 647) at 635 nm. Both lasers are operated at 40 MHz repetition rate. The laser light is reflected towards the microscope by a multiband dichroic mirror (ZT 405/488/561/640rpc, Chroma). The excitation intensity of the laser is kept at 20 µW (measured at the microscope entrance port) during the whole measurements. The excitation light is focused on a single ZMW by a Zeiss C-Apochromat 63x, 1.2 NA water immersion objective lens. The fluorescence light arising from the sample is collected by the same objective in an epifluorescence configuration and is passed through the multiband dichroic (ZT 405/488/561/640rpc, Chroma) and an emission filter (ZET405/488/565/640mv2, Chroma) to separate the fluorescence from the back reflected laser light. The green and red fluorescence signals are separated by a dichroic mirror (ZT633RDC, Chroma). Both detection channels are equipped with a 50 µm pinhole and emission filters to further spatially and spectrally filter the fluorescence light (green ET570LP and ET595/50m Chroma filters, red 640LP and 655LP Chroma filters). Two single photon avalanche photodiodes (MPD-5CTC with < 50 ps timing jitter, Picoquant) are used to detect the green and red fluorescence photons. Each detection event is tagged with its individual time and channel information in a time-tagged-time resolved (TTTR) mode by a fast time correlated single photon counting module (HydraHarp 400, Picoquant). The temporal resolutions of the fluorescence measurements are 38 ps and 110 ps upon 557 nm and 635 nm excitation respectively. All the measurements presented here were recorded 5 minutes after the solution containing the DNA molecules was spread over the ZMW surface. No visible change was detected during two hours of incubation of the ZMW with the DNA. We conclude that the observed behavior due to binding/unbinding (and/or bleaching) is in a stationary temporal regime. While we typically present time traces of 10 to 15 s duration in our figures, the total acquisition to record the FCS and lifetime traces is 60 s.

**Fluorescence correlation spectroscopy (FCS) data analysis**. FCS traces are obtained from the autocorrelation of the fluorescence intensity time trace. The resulting FCS traces are fitted using a 3-dimensional Brownian diffusion model with an additional blinking term:[65]

$$G(\tau) = \frac{1}{N}\left[\left(1 + \frac{T_{ds}}{1-T_{ds}}exp\left(-\frac{\tau}{\tau_{ds}}\right)\right)\right]\left[\sum_i \alpha_i \left(1 + \frac{\tau}{\tau_{D,i}}\right)^{-1}\left(1 + \frac{\tau}{\kappa^2 \tau_{d,i}}\right)^{-0.5}\right] \quad (1)$$

Where N is the total number of molecules in the observation volume, $T_{ds}$ is the fraction of the dyes in the dark state, $\tau_{ds}$ is the lifetime of the dark state, $\alpha_i$ is the fraction of the populations with mean diffusion time $\tau_{D,i}$ through the observation volume and $\kappa$ is the structure parameter of the ZMW observation volume representing the ratio between the axial and transverse dimensions of the ZMW observation volume. The above model equation was found to empirically describe well the FCS traces inside ZMWs (Figure 2 and 4), provided that the structure parameter $\kappa = 1$ as found previously. In the absence of sticking, a 1-component diffusion model (i =1) accounts well for the experimental data. However, in the presence of sticking to the ZMW surface, an additional component is clearly visible on the FCS correlation functions at long lag times. Therefore, for these cases we use a 2-species model in the above equation (i =2) to distinguish between the contributions of the fast and slow FCS components.

**Lifetime analysis.** The time correlated single photon counting (TCSPC) histograms are fitted by a Levenberg-Marquard optimization performed on the SymPhoTime 64 software (PicoQuant GmbH). An iterative reconvolution fit is carried out by taking into account the instrument response function (IRF). The time gate for fitting the TCSPC histograms are set to ensure that there are always 95% of photons in the region of interest. All the TCSPC histograms obtained for the dye-DNA constructs in the confocal measurements are fitted with a single exponential function. The fluorescence lifetime values of the Alexa 546, Alexa 647, Atto 550 and Atto 647N dye labelled DNA in the confocal measurements are 3.5 ns, 1.4 ns, 3.4 ns and 4.0 ns respectively. The TCSPC histograms obtained for the Atto-DNA constructs in the presence of untreated ZMW and PEG1000 passivated ZMW are fitted into a triexponential and biexponential function respectively, in order to provide a better fit to the data. In the case of multiexponential fit the intensity weighted average lifetime is used to represent the fluorescence lifetime. The intensity weighted average lifetime for the Atto 550 and Atto 647N labelled DNA in the presence of uncoated ZMW is found to be 1.02 ns and 1.41 ns respectively. A PEG-1000 passivation of the ZMW leads to an increase in the intensity weighted average lifetime value to 1.39 and 1.74 ns for Atto 550 and Atto 647N labelled DNA constructs respectively. The Alexa labelled DNA constructs exhibit similar TCSPC histograms in the presence of both passivated and unpassivated ZMW. A bi and triexponetial fit is found to be a satisfactory approximation to the TCSPC histograms obtained for Alexa 546-DNA and Alexa 647-DNA constructs respectively in the presence of ZMW. The fluorescence lifetime determined for Alexa 546 and Alexa 647 labelled DNA inside ZMW is 1.78 ns and 0.68 ns respectively. For Alexa 546 and Alexa 647, an extremely fast component (~5 ps) is detected in the TCSPC histogram in the presence of ZMW due to metal induced backreflection of the laser light. This extremely fast lifetime component is excluded from the analysis of the fluorescence lifetime for Alexa dyes inside ZMW.

**Acknowledgments**

The authors thank Antonin Moreau and Julien Lumeau for help with the aluminum deposition. This project has received funding from the European Research Council (ERC) under the European Union's Horizon 2020 research and innovation programme (grant agreement 723241) and from the Agence Nationale de la Recherche (ANR) under grant agreement ANR-17-CE09-0026-01.

**Author Contributions**

S. P., M. B. and J.W. conceived and designed the experiments; S. P. and M. B. carried out the experiments and data analysis; J.-B. C. prepared the ZMW samples; S.P. and J.W. wrote the manuscript; all authors discussed the results and commented on the manuscript.

**Additional Information**

**Competing financial interests:** The authors declare no competing financial interests.

**Supporting Information for**

**Surface passivation of zero-mode waveguide nanostructures:**

**benchmarking protocols and fluorescent labels**

Satyajit Patra, Mikhail Baibakov, Jean-Benoît Claude, Jérôme Wenger*

*Aix Marseille Univ, CNRS, Centrale Marseille, Institut Fresnel, 13013 Marseille, France*

*\* Corresponding author: [jerome.wenger@fresnel.fr](jerome.wenger@fresnel.fr)*

**Contents**:

1. Fluorescence intensity time traces of green dyes-DNA conjugates inside passivated and unpassivated ZMWs (Figure S1).

2. Apparent fluorescence brightness per molecule enhancement inside ZMWs for different dye-DNA conjugates (Figure S2).

3. Fluorescence lifetime plot of Alexa 546 and Atto 550 labelled DNA in confocal and inside ZMW (Figure S3).

4. FCS correlation functions for different PEG coatings (Figure S4).

## S1. Fluorescence intensity time traces of green dyes-DNA conjugates

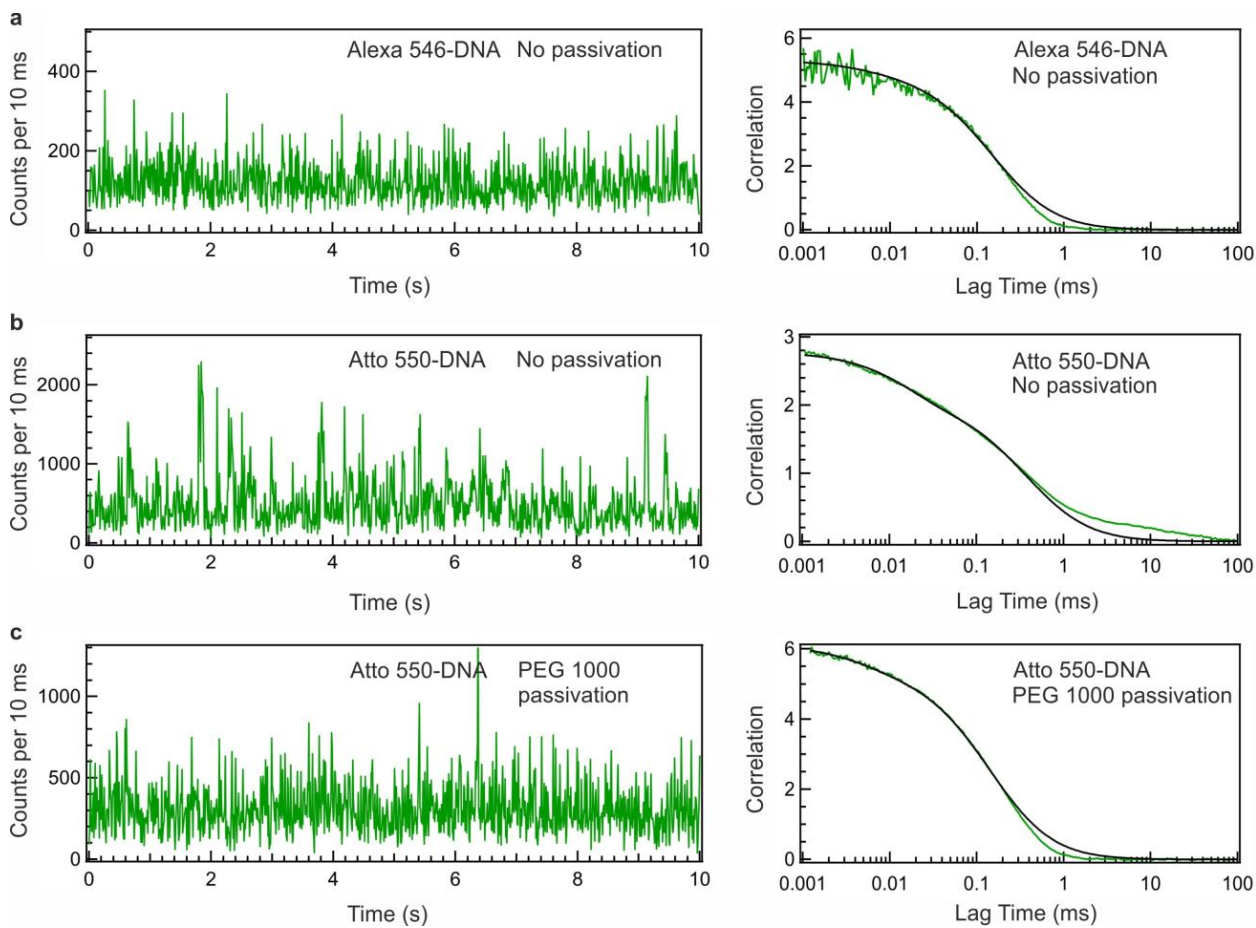

**Figure S1**: Results obtained from the diffusion of the DNA labelled with green dyes through the attoliter ($10^{-18}$L) ZMW observation volume. (a) Fluorescence intensity vs time and the corresponding FCS trace obtained for the diffusion of the Alexa 546-DNA through the observation volume of unpassivated ZMW. (b) The fluorescence time trace for Atto 550-DNA inside an unpassivated ZMW shows spikes indicating sticking of the DNA on the ZMW surface, as confirmed by the long diffusing component on the FCS data. (c) Passivation of the ZMW surface with PEG 1000 efficiently eliminates the sticking of Atto 550 labelled DNA sample on the ZMW. The DNA concentration is 100 nM and the ZMW diameter is 110 nm.

**S2. Apparent fluorescence brightness per molecule enhancement inside ZMWs for different dye-DNA conjugates**

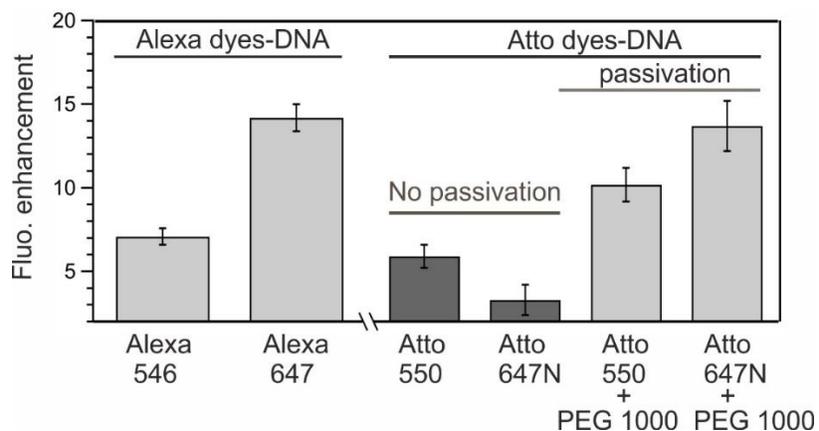

**Figure S2**: Enhancement factors for the fluorescence brightness per molecule inside 110 nm aluminum ZMWs for different dye-DNA conjugates. The brightness per molecule is computed as the ratio of the average fluorescence intensity divided by the number of molecules estimated by FCS analysis (Fig. 2e), then the enhancement factor corresponds to the ratio of the brightness per molecule in the ZMW to its reference value in the confocal setup. The fluorescence enhancement factors for the Alexa dye labelled DNAs are identical inside passivated and unpassivated ZMWs. The Atto-DNA constructs exhibits higher apparent fluorescence enhancement when the nanoapertures are passivated with PEG 1000 as a result of the misestimate of the number of molecules seen in Fig. 2e when surface sticking occurs.

## S3. Fluorescence lifetime plot of Alexa 546 and Atto 550 labelled DNA

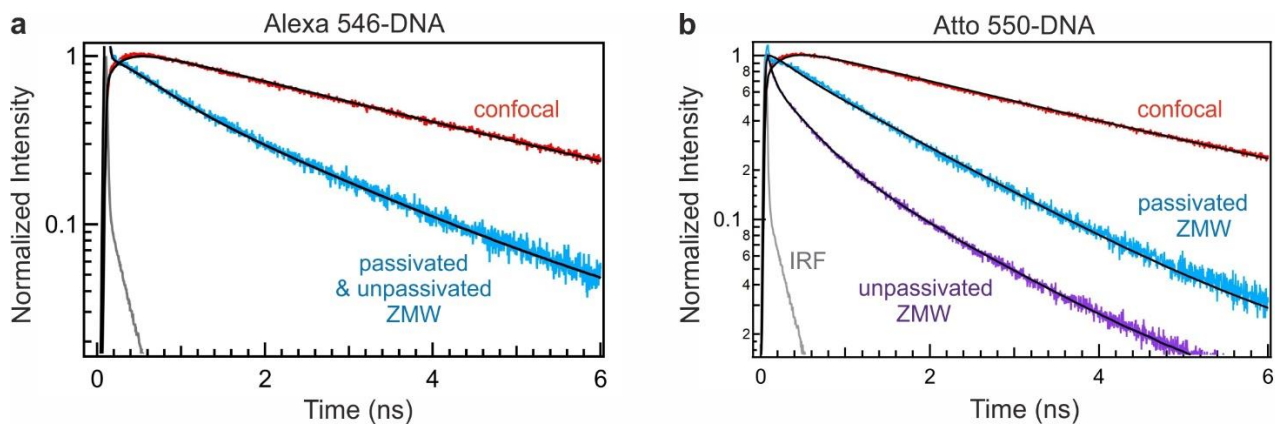

**Figure S3**: Fluorescence intensity decay plots of Alexa 546 and Atto 550 labelled DNA in the confocal and in the presence of a 110 nm ZMW. For Alexa 546-DNA the results are identical for passivated and unpassivated ZMWs. For Atto 550-DNA, we find a shorter fluorescence lifetime inside unpassivated ZMW than the passivated one, confirming the occurrence of surface sticking inside unpassivated ZMWs.

## S4. FCS correlation functions for different PEG coatings

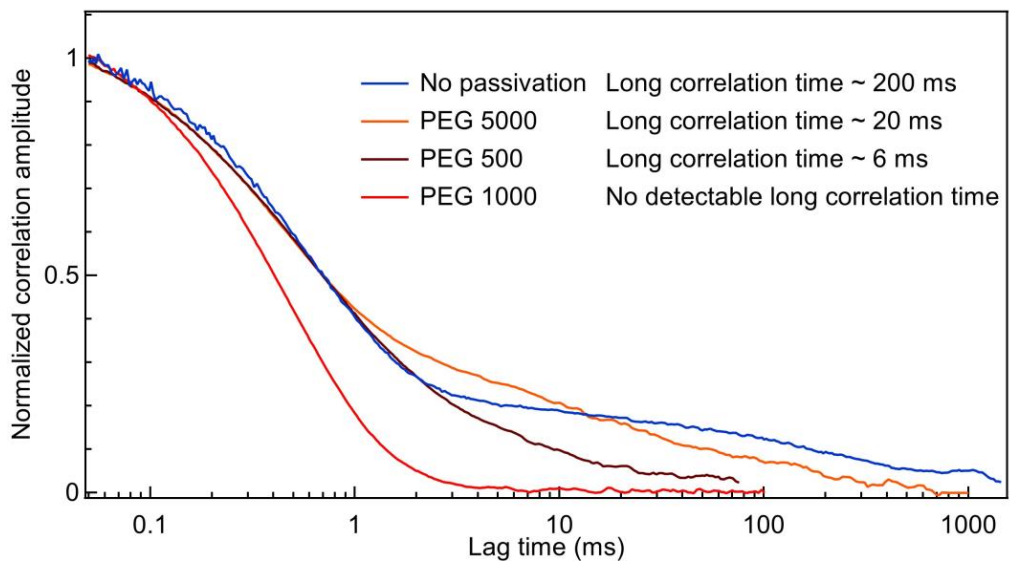

**Figure S4**: Comparison of normalized FCS correlation functions for different PEG coatings and untreated ZMW surface, in the case of Atto 647N-DNA conjugates in a 110 nm diameter aluminum ZMW (similar to Fig. 4 of the main document). The presence of the long correlation component indicates Atto647N-DNA binding and unbinding (or bleaching) from the ZMW surface. The presence of the PEG coating reduces this long correlation time significantly, which demonstrates that the surface is passivated although sometimes not completely (PEG 500, PEG 5000).